\documentclass[11pt,aps,onecolumn,prl,showpacs,notitlepage]{revtex4-1}
\usepackage{graphicx}
\usepackage{color}
\usepackage{amsmath,amsthm,amssymb}
\usepackage[bf,Large]{}

\begin{document}
\title{A Minimal tight-binding model for ferromagnetic canted bilayer manganites}
\author{M. Baublitz\textsuperscript{1,2}, C. Lane\textsuperscript{1*}, Hsin Lin\textsuperscript{1,3}, Hasnain Hafiz\textsuperscript{1}, R.S. Markiewicz\textsuperscript{1}, B. Barbiellini\textsuperscript{1}, Z. Sun\textsuperscript{4}, D.S. Dessau\textsuperscript{4} \& A. Bansil\textsuperscript{1}}
\affiliation{
\textsuperscript{1}Physics Department, Northeastern University, Boston MA 02115, USA, \\
\textsuperscript{2}College of General Studies, Boston University, Boston MA 02215, USA, \\
\textsuperscript{3}Graphene Research Centre and Department of Physics, National University of Singapore, Singapore 117542, \\
\textsuperscript{4}Department of Physics, University of Colorado, Boulder CO 80309, USA\\
\textsuperscript{*}c.lane@neu.edu}
\date{version of \today}
\begin{abstract}
Half-metallicity in materials has been a subject of extensive research due to its potential for applications in spintronics. Ferromagnetic manganites have been seen as a good candidate, and aside from a small minority-spin pocket observed in La$_{2-2x}$Sr$_{1+2x}$Mn$_{2}$O$_{7}$ $(x=0.38)$, transport measurements show that ferromagnetic manganites essentially behave like half metals. Here we develop robust tight-binding models to describe the electronic band structure of the majority as well as minority spin states of ferromagnetic, spin-canted antiferromagnetic, and fully antiferromagnetic bilayer manganites. Both the bilayer coupling between the MnO$_2$ planes and the mixing of the $|x^2 - y^2>$ and $|3z^2 - r^2>$ Mn 3d orbitals play an important role in the subtle behavior of the bilayer splitting. Effects of $k_z$ dispersion are included.

\end{abstract}
\maketitle
Manganites\textsuperscript{1,2} have been widely studied because of their remarkable properties of colossal magnetoresistance\textsuperscript{3} and possible half-metallicity\textsuperscript{4,5}, where electrons of one spin are metallic and those of the opposite spin are insulating. Metals with a high degree of spin polarization at the Fermi level are of great interest for possible applications in spintronics\textsuperscript{6,7}, enabling the processing of data and memory storage via spins instead of conventional methods involving transport of charge.

Manganites are quasi-two dimensional materials with layered structures similar to those of high $T_c$ cuprate superconductors. The structure of LaSr$_2$Mn$_2$O$_ 7$ (LSMO) resembles that of the prototypical perovskite mineral CaTiO$_3$, and it can be described in terms of a stacking of double layers\textsuperscript{8} of interconnected MnO$_6$ octahedra in which Mn atoms sit at the center and oxygen atoms occupy corners of the octahedron. The MnO$_6$ octahedra are distorted, and a crystal-field-splitting parameter $E_z$ can be used to characterize the splitting between the $|x^2 - y^2>$ and the $|3z^2 - r^2>$ 3d levels of the Mn atoms.\textsuperscript{9}

Doped bilayer manganites display a rich phase diagram, which includes
a ferromagnetic (FM) phase as well as a more subtle antiferromagnetic (AFM)
state where spins are aligned ferromagnetically within the MnO planes, 
but canted antiferromagnetically between 
the adjacent MnO planes\textsuperscript{10}. 
The bilayer coupling plays a key role 
in stabilizing the FM phase by preserving
phase coherence between the neighboring MnO planes. 
When doping with Sr from 
$x=0.38$ to $0.59$, where $x$ is the electronic doping away from half-filling or, equivalently, the ratio of Sr to La, 
strength of the bilayer coupling decreases due to the canting of spins between the adjacent layers,
and finally it vanishes in the fully AFM phase. 
Angle-resolved photoemission spectroscopy (ARPES) 
experiments\textsuperscript{11} 
show that the ferromagnetic compound 
($x=0.38$) exhibits a finite bilayer splitting due to interlayer hopping,
while the antiferromagnetic compound ($x=0.59$) 
has zero bilayer splitting since the adjacent layers are oppositely spin polarized.

A ferromagnetic calculation on LSMO 
based on the generalized gradient approximation (GGA)\textsuperscript{11}
shows that bands at the Fermi energy ($E_F$) 
are primarily of e$_g$ character\textsuperscript{12} 
(i.e. Mn 3d $|x^2 - y^2>$ and $|3z^2 - r^2>$)
for the majority spins, and of t$_{2g}$ character (i.e. Mn 3d $|xy>$) 
for the minority spins, 
which is consistent with ARPES 
results\textsuperscript{13}.
Previous comparisons between ARPES
and density functional theory (DFT) computations have revealed
that the GGA gives a better description than the 
local spin density approximation
(LSDA), and that the LSDA corrected by a Hubbard parameter (LSDA+U)
gives an even poorer description of the ARPES data\textsuperscript{11}.
The GGA provides a simple but potentially accurate step beyond LSDA
which can improve the description of magnetic properties of 
the 3d electronic shell\textsuperscript{14}.

The metallic conductivity in the FM phase 
can be explained within 
the double-exchange (DE) mechanism\textsuperscript{15}, 
where e$_g$ electrons hop between the Mn sites 
through hybridization with the oxygen 2p orbitals. 
While the DE mechanism appears to capture 
the tendency towards ferromagnetism, the 
oxygen orbitals must be explicitly included 
to explain correctly the metal insulator 
transition at the Curie temperature\textsuperscript{16}.

Since the DFT band structure is found to be that 
of a nearly half-metallic ferromagnet with a small minority-spin FS (Fermi surface), 
most studies in the literature focus only on the majority bands described within
simple tight-binding (TB) models\textsuperscript{17}, 
neglecting the minority bands.
Here, we present a more realistic yet transparent TB model 
which incorporates the bonding and antibonding 
$|x^2 - y^2>$ as well as the $|3z^2 - r^2>$ orbitals, including 
the minority states as observed via ARPES in 
the FM\textsuperscript{13} and AFM\textsuperscript{11} states.
Recall that in the cuprates there is strong copper-oxygen hybridization, but if one is mainly interested in the antibonding band near the Fermi level, one can study an effective, copper-only model.  In this spirit, we develop an effective Mn-only model here, which includes the minority bands in order to provide a precise description of the minority electrons in determining the spin polarization at the Fermi level, a key ingredient needed for the design of spintronics devices. 
We delineate how our model Hamiltonian gives insight into  
the delicate interplay between the effects of orbital mixing and nesting features, which impact the static susceptibility 
and drive exotic phase transitions\textsuperscript{18}.
Our approach can also allow a precise determination of the 
occupancy of the minority $t_{2g}$ electrons 
through an analysis of the experimental FSs.

\section{Results}
{\bf Band character near E$_F$}  
In the DFT-based band structure, E$_F$ cuts through the 
majority $|x^2 - y^2>$ and $|3z^2 - r^2>$ bands, 
while there are only small electron pockets in 
the minority $|xy>$ bands.   
Coupling between the two MnO layers in the FM state produces 
bonding and antibonding bands, which are 
directly observed in 
experiments\textsuperscript{19}. Accordingly, our fitting procedure is based on a combination of four majority and two minority bands in order to accurately capture the near-E$_F$ physics of the system.

For the majority $e_{g}$ bands, the strength of bilayer coupling for $|x^2 - y^2>$ orbitals is much weaker than that for $|3z^2 - r^2>$ orbitals because the lobes of $|x^2 - y^2>$ orbitals lie in-plane, while those of $|3z^2 - r^2>$ orbitals point out-of-the-plane. The bilayer coupling of various orbitals without hybridization can be seen along the $\Gamma(0,0)$-$X(\pi,\pi)$ line 
in Figure 1, where the two $|x^2 - y^2>$ bands  
are nearly degenerate and the two $|3z^2 - r^2>$ bands are
split with a separation of $\approx$ 1.1 eV. Away from the nodal direction, the 
$|x^2 - y^2>$ and $|3z^2 - r^2>$ orbitals hybridize, and the splitting of the related bands becomes more complex. Near the $M(\pi,0)$ point, the two lowest bands are primarily of $|x^2 - y^2>$ character. The mixing with $|3z^2 - r^2>$ increases the splitting to $\approx$ 250 meV.

Regarding the t$_{2g}$ minority bands,
since the lobes of $|xy>$ orbitals lie in-plane, 
strength of the bilayer coupling is small. 
Unlike $|x^2 - y^2>$, the 
lobes of $|xy>$ are rotated 45$^{\circ}$ from the MnO direction, 
so that the hybridization 
with other bands and the resulting splittings reach 
their maximum value at the $X$-point.

{\bf Tight-binding Model: Majority Spin} 
Since there is a large
exchange splitting, we discuss the
majority and minority bands separately.
This section presents the TB model for the majority
spins, obtained by fitting to the first
principles band structure. 
The four bands near $E_F$ are predominantly associated with the $eg$ orbitals of Mn 3d,
$|x^2 - y^2>$ and $|3z^2 - r^2>$, so that the minimal TB model involves four orbitals per primitive unit cell.
In this connection, it is useful to proceed in steps, and accordingly, we first discuss a 2-dimensional (2D) model with bilayer splitting, 
followed by the inclusion of effects of $k_{z}$-dispersion.

For the 2D model,
the relevant symmetric (+) and antisymmetric (-)
combinations of the orbitals
decouple, and the 4 $\times$ 4 Hamiltonian
reduces to two 2 $\times$ 2 Hamiltonians,
$H_{\pm}$, where the basis functions are
$\psi_{1\pm}$ and $\psi_{2\pm}$ with the
subscripts 1 and 2 referring to the $|x^2
- y^2>$ and $|3z^2 - r^2>$ orbitals,
respectively. 
The Hamiltonian matrices are
\begin{align}
H_{\pm}&=\left(\begin{matrix} 3H_{11}+E_z/2\pm t_{bi1}&{\sqrt 3}H_{12} \\ {\sqrt 3}H_{12} & H_{22} -E_z/2\pm H_{bi2}\cr \end{matrix}\right),
\end{align}
 where
\begin{align}
H_{11}&=t_{11}(c_x(a)+c_y(a))/2+t'_{11}c_x(a)c_y(a)+t''_{11}(c_x(2a)+c_y(2a))/2+t'''_{11}(c_x(3a)+c_y(3a))/2,\nonumber\\
H_{22}&=t_{22}(c_x(a)+c_y(a))/2+t'_{22}c_x(a)c_y(a),\\
H_{12}&=t_{12}(c_x(a)-c_y(a))/2+t'_{12}(c_x(2a)-c_y(2a))/2+t''_{12}(c_x(2a)c_y(a)-c_x(a)c_y(2a)),\nonumber
\end{align}
$c_i(\alpha a)=cos(k_i\alpha a)$, $i=x,y$, and $\alpha$ is an integer.
$t_{ij}$ are the hopping parameters where $t_{11}$ is the hopping between the $|x^2
- y^2>$ orbitals, $t_{22}$ for the $|3z^2 - r^2>$ orbitals, and $t_{12}$ between the $|x^2- y^2>$ and $|3z^2 - r^2>$ orbitals.
Here the nearest neighbor hopping is denoted by $t_{ij}$, the next nearest hopping by $t'_{ij}$, and the higher order hoppings are denoted by a larger number of primes as superscripts.
Note that the two
matrices in Eq. 1 are identical except for the
last term on the main diagonal, differing only in the sign of the
bilayer hopping terms $t_{bi1}$ and $H_{bi2}=t_{bi2}+t'_{bi2}(c_x(a)+c_y(a))/2$. The chemical potential $\mu$ is obtained via a least squares fit to the first-principles GGA bands.

If the hopping parameters are deduced within the
Slater-Koster model\textsuperscript{20}, one would obtain   
$t_{11}=t_{22}=t_{12}=t_{bi2}$, 
and $t'_{11} = t'_{22}$. However, we found an
improved fit by letting the parameters
deviate from these constraints. A number of
additional hopping terms were tested, but
found to give negligible improvements
and discarded. A least squares minimization program was
used to obtain the optimized TB 
parameters, which are listed in Table 1 (2D model). 

Values of TB parameters in Table 1 are consistent with previous results on cubic manganites\textsuperscript{17}. 
It is reasonable that the four nearest neighbor parameters ($t_{11}$, $t_{22}$, $t_{12}$, and $t_{bi2}$) are the largest in absolute magnitude and are
the most important fitting parameters. Sign differences between $t'_{11}$, $t'_{22}$ and $t'_{12}$ control the presence of a closed FS related to $|3z^2 - r^2>$ bands and an open FS from $|x^2 - y^2>$ bands, consistent with earlier studies\textsuperscript{18}. TB
parameters with small
magnitudes ($t''$, and $t'''$)
involve overlap between
more distant neighbors. We emphasize that even though $t''$ and $t'''$ are small, they contribute significantly to the overall goodness of the fit.
A small value of $t_{bi1}$ reflects weak intra-layer interactions between the $|x^2 - y^2>$ orbitals 
due to the orientation of these orbitals. 
Since the magnitude of the crystal field splitting parameter $E_z$ is smaller 
than that of $t_{12}$, the hybridization of $|x^2 - y^2>$ and $|3z^2 - r^2>$ is 
significant when $H_{12}$ is nonzero.

Figure 2 compares the model
TB bands (open circles)
with the corresponding DFT results (solid
dots). While the full 2D model is considered in Figure 2a, we also show in 
Figure 2(b), results of a much
simpler TB model that employs
only two parameters ($E_z$ and $t$) with 
$t_{11}=t_{22}=t_{12}=t_{bi2}$.
For the simple model of Figure 2b, the parameter values 
($t =-0.431$ eV, $E_z=-0.057$ eV, and $\mu=0.616$ eV) 
were obtained via an optimal fit to the first-principles bands.
It is obvious that the 2D TB model results
shown in Figure 2a provide a vastly
improved fit compared to
the simple two parameter model in Figure 2b.
The agreement in Figure 2a between the
TB model and the first principles
calculations is overall very good and 
the TB model correctly reproduces salient features
of the band structure. 

At $\Gamma$, the two lowest
energy bands are found to be nearly degenerate
in both the TB model and the first principles
calculations, with a splitting of $-2t_{bi1}=0.044$eV in the TB model. Following these two bands along $\Gamma-X$,
one finds that the two larger dispersing bands with $|x^2 - y^2>$
character have small bilayer splitting due to the small value of $t_{bi1}$.
The two other bands in the same direction are
of $|3z^2 - r^2>$ character, and
exhibit a larger bilayer splitting of $-2H_{bi2}=1.09$eV.
Because $H_{bi2}$ contains the next-nearest-neighbor hopping terms, the bilayer splitting of $|3z^2 - r^2>$ bands develops an in-plane k-dependence. 
As a result, dispersion of the antibonding band is larger than 
that of the bonding band.
Along the $\Gamma-M$ and $X-M$ directions, $H_{12}$ is non-zero,
leading to the mixing of $|x^2 - y^2>$ and $|3z^2 - r^2>$ bands.
At the $M$-point, $H_{12}$ reaches its maximum value, yielding a 
complex bilayer splitting of the Van Hove singularities.
In other words, the bare bilayer splitting of $|x^2 - y^2>$ is $\approx$
50 meV, but hybridization with $|3z^2 - r^2>$ enhances this splitting to $\approx$ 290 meV near $M$ in the TB model as follows:
\begin{align}
\Delta E&=-(t_{bi1}+H_{bi2})-S_{-}+S_{+}
\end{align}
where $S_{\pm}=\sqrt{(C_{\pm}\mp H_{bi2})^{2}+12H_{12}^{2}}$ and $C_{\pm}=3H_{11}-H_{22}+E_{z}\pm t_{bi1}$.

Figure 3 compares the 2D-TB (open circles) and first-principles (dots) FSs. Agreement is seen to be quite good. 
The three pieces of FS are labeled by `1', `2' and `3'.
The larger squarish pocket `1'
centered at $X$ is a mix of $|x^2 - y^2>$ and $|3z^2 - r^2>$,
the smaller squarish pocket `3' around the $\Gamma$-point is primarily of 
$|3z^2 - r^2>$ character, and  
the rounded FS `2' lying between `1' and `3' centered at $X$ 
is mostly of $|x^2 - y^2>$ character.
For comparison Figure 3b shows the FS from the
simple two parameter TB model of 
Figure 2b, and we see again that this simple model gives a poor representation of the actual FS.

Recall that in the cuprates, there is a small but finite $k_z$-dispersion\textsuperscript{21-24},
which is also the case in the manganites.  
Since the $|3z^2 - r^2>$ orbitals have lobes pointing out of the plane, 
the interlayer hoppings are associated with $|3z^2 - r^2>$ bands. 
In the 3D model, the 4 $\times$ 4 Hamiltonian now cannot be 
reduced to two 2 $\times$ 2 Hamiltonians because of the body-centered crystal structure.  The basis functions are $|x^2 - y^2>$ and $|3z^2 - r^2>$ for the upper and lower MnO$_2$ layers. By including interlayer hopping $t_z$ between $|3z^2 - r^2>$ orbitals and the intra-layer hopping $t'_z$ for $|3z^2 - r^2>$ orbitals, we obtain the 
Hamiltonian matrix:
\setlength{\arraycolsep}{-1pt}
\begin{align}
H_4&=\nonumber\\
&\left(\begin{matrix} 3H_{11}+E_z/2 &{\sqrt 3}H_{12} & t_{bi1} & 0 \cr
{\sqrt 3}H_{12} & H_{22}-E_z/2+2t'_zc_x(\frac{a}{2})c_y(\frac{a}{2})c_z(\frac{c}{2}) & 0 & H_{bi2}+t_zc_x(\frac{a}{2})c_y(\frac{c}{2})\exp(i\frac{k_zc}{2})\cr
 t_{bi1} & 0 &3H_{11}+E_z/2 &{\sqrt 3}H_{12}  \cr
0 & H_{bi2}+t_zc_x(\frac{a}{2})c_y(\frac{a}{2})\exp(-i\frac{k_zc}{2})&{\sqrt 3}H_{12} & H_{22}-E_z/2+2t'_zc_x(\frac{a}{2})c_y(\frac{a}{2})c_z(\frac{c}{2}) & \cr\end{matrix}
\right)
\end{align}

where $c_z(c)=cos(k_zc)$ and c is the lattice constant in the z-direction, which is approximately 5 times larger than the in-plane lattice constant a. The parameters obtained by fitting to the DFT bands are listed 
in Table 1 (3D model). Compared to the 2D model, the bilayer hopping parameters $t_{bi1}$, $t_{bi2}$ and 
$t'_{bi2}$ are significantly modified. $t_{22}$ and $E_z$ change by about 30meV while other terms undergo only slight modifications. Plausible values of parameters are retained in the 3D model.

The effect of $k_z$-dispersion in the 3D model can be seen by 
comparing the FSs at $k_zc=0$ and $k_zc=2\pi$ 
as shown in Figure 4. While FS `2' with 
mostly $|x^2 - y^2>$ character remains unchanged, the FS piece `3' with primarily $|3z^2 - r^2>$ character changes significantly.
`3' is squarish at $k_zc=0$ but 
becomes smaller and rounded at $k_zc=2\pi$ (`3$^{\prime}$').
Although `1' contains a significant $|3z^2 - r^2>$ contribution, 
the effect of $k_z$-dispersion on this FS piece is much smaller than on `3'. 
`1' and `1$^{\prime}$' match when $k_x a=\pi$ or $k_y a=\pi$ because 
the interlayer hopping terms $t_z$ and $t'_z$ have zero contribution due to the $c_x({\frac 1 2}a)c_y({\frac 1 2}a)$ dependence in the body-centered structure. `1' and `1$^{\prime}$' almost match when $k_x a=k_y a$ because $t_{bi1}$ is almost 
zero. Thus `1' and `1$^{\prime}$' can differ only away from the high symmetry k-points and 
this piece of the FS is cylinder-like in 3D.

{\bf Tight-binding Model: Minority
Spin} Due to the large exchange splitting, we only need to consider two bands in the case of minority spins, which are associated with the $t_{2g}$ $|xy>$ orbitals of the upper and lower MnO$_{2}$ layers.
 The 2 $\times$ 2 model
Hamiltonian 
given below is diagonal with a bilayer splitting of $\Delta$ between the upper and lower $|xy>$ bands.
\begin{align}
H&=\left(\begin{matrix}H_{11}+\Delta/2& 0 \cr
0& H_{22}
-\Delta/2\cr\end{matrix}\right)
\end{align}
where 
\begin{align}
H_{11}&=t_{11} (c_x(a)+c_y(a))+t'_{11} c_x(a)c_y(a)+t''_{11}(c_x(2a)+c_y(2a))+t'''_{11} c_x(2a)c_y(2a),\\
H_{22}&=t_{22} (c_x(a)+c_y(a))+t'_{22} c_x(a)c_y(a)+t''_{22}(c_x(2a)+c_y(2a))+t'''_{22} c_x(2a)c_y(2a).\nonumber
\end{align}

Table 2 lists the parameters obtained from fitting first-principles band structure. Figure 5 compares the
parameterized TB
bands (open circles) with the first-principles GGA bands (solid
dots).
The minority spin FSs are overlayed
in Figure 3 as triangles, and form two small pockets around $\Gamma$, as observed also in the ARPES experiments\textsuperscript{13}.


{\bf Doping and Magnetic Structure.} We now turn to discuss how the low-energy electronic structure of the rich variety of magnetic phases displayed by LSMO is captured by our 2D and 3D TB models. Kubota {\em et al.}\textsuperscript{10} have shown that the magnetic structure of LSMO is intimately connected with doping, and that it can be parameterized in terms of $\theta_{cant}$, the spin canting angle between the neighboring FM planes.  The behavior of $\theta_{cant}$, deduced from experiments, shows a FM structure ($\theta_{cant}=0^{\circ}$) for $0.32\leq x \leq 0.38$, with the value of $\theta_{cant}$ becoming finite at $x\approx 0.39$, and reaching $180^{\circ}$ for $x\geq 0.48$\textsuperscript{10}. In the 2D and 3D models discussed above, for doping greater than $x=0.38$, the value of E$_F$ was found by assuming a rigid band type approximation\textsuperscript{25} where the total number of occupied electrons $N$ is given by $N=2(1-x)$ at doping $x$.
Over this doping range the exchange splitting from GGA was taken to be constant since the spins are ferromagnetically aligned in planes and the in-plane lattice parameters are not sensitive to doping\textsuperscript{10}.
We then invoke the argument of Anderson et al.\textsuperscript{26} that the transfer integral between any two ions depends on $\cos(\theta /2)$ where $\theta$ is the angle between their spins
on neighboring layers as the magnetic state changes from FM to AFM. 
We thus replaced the bilayer TB parameters $H_{bi2}$, $t_{bi1}$ and $\Delta$ by $\cos(\theta_{cant}/2)H_{bi2}$, $\cos(\theta_{cant}/2)t_{bi1}$ and $\cos(\theta_{cant}/2)\Delta$, and for the 3D model $t_{z}$ was also replaced with $\cos(\theta_{cant}/2)t_{z}$, using the experimental values of $\theta_{cant}$ at the corresponding dopings given by Kubota et al.\textsuperscript{10}.

Table 3 gives values of $\Delta E_{F}$ (where $\Delta E_{F}$ is measured with respect to E$_{F}$ at x=0.50 in the FM state), number of minority electrons, $\Delta n$, number of majority electrons, $1-x-\Delta n$, total number of electrons, $1-x$, canting angle, $\theta_{cant}$, and the magnetic moment $\mu_{B}$, all per Mn atom for the doping range $0.38$-$0.59$, as obtained within our 2D and 3D models. Table 4 provides the same quantities over this doping range only in the FM state appropriate for saturating magnetic fields. [The doping range used for calculations in Tables 3 and 4 does not include the experimentally observed anomalous FS behavior\textsuperscript{27}.] The magnetic moment $\mu_{B}$ per Mn atom, including the contribution of the three occupied $t_{2g}$ orbitals, is given by $\mu_{B}=1-x-2\Delta n+3$, and its values are consistent with magnetic Compton experiments\textsuperscript{28,29}. The number of minority electrons, $\Delta n$, found in recent ARPES experiments\textsuperscript{13} 
is also in good agreement with the corresponding values in Table 3. We find that, in comparison to the GGA, the LSDA underestimates the exchange splitting by 20\% and thus overestimates the number of minority. On the other hand, the TB parameters based on LSDA and GGA band structures differ only within 1\%.

Figure 6a compares the experimental FS for $x=0.38$ (FM)\textsuperscript{13}, with the corresponding 2D TB model predictions. Good agreement is seen between theory and experiment for the FS pieces related to the $d_{3z^2 - r^2}$ (red line), the anti-bonding $d_{x^2 - y^2}$ (green line), and the minority pockets (pink and black lines). The bonding hole-pocket (blue) is invisible at this photon energy due to matrix element effects\textsuperscript{13,19,21,22}. 
In order to account for the coexistence of metallic and nonmetallic regions for $x\leq 0.38$, which has been interpreted as arising from a phase separation into hole-rich and hole-poor regions\textsuperscript{27}, we found it necessary to adjust the doping of the theoretical FS at x=0.38 to an effective dopping of x=0.43.
 Figure 6b shows the $x=0.59$\textsuperscript{11} experimental AFM FS, along with the corresponding 2D TB model results. Here also we find good agreement for the bonding and anti-bonding $d_{x^2 - y^2}$ bands (blue and green lines). The same level of agreement between theory and experiment is also found for the 
3D model, which is to be expected since the values in Tables 3 and 4 for the 
2D and 3D models are very similar. 

\section{Discussion} 

The double-layered manganites, La$_{2-2x}$ Sr$_{1+2x}$Mn$_{2}$O$_{7}$, have attracted much attention in recent years as model systems that present a wide range of transport and magnetic properties as a function of temperature, doping and magnetic field. In the FM phase at $x=0.38$, the majority $t_{2g}$ electrons of Mn lie well below the Fermi level and are thus quite inert. Therefore, 
key to the understanding of the manganites is the behavior of the Mn magnetic 
electrons with $e_g$ character ($|x^2 - y^2>$ and $|3z^2 - r^2>$). The results of magnetic Compton experiments\textsuperscript{28,28} reveal that the FM order weakens when the occupation of the  $|3z^2 - r^2>$  majority state decreases.
For spintronics applications, it is important to note that the Fermi level 
in the FM phase lies slightly above the bottom of the minority-spin conduction band, yielding a nearly half-metallic ferromagnet. The unwanted FS-pocket can be reduced in volume by increasing the doping $x$. However, the Mn spins (aligned ferromagnetically within the MnO planes) become canted antiferromagnetically between the adjacent MnO planes as $x$ increases,  
leading to a competing AFM order which destroys the FM phase.

In order to understand this interesting phenomenology, we have developed 
a TB model encompassing both the FM and AFM phases, which correctly captures the low-energy electronic structure of LSMO using a minimal basis set. The complex bilayer splitting in the majority spins is well reproduced.
In particular, the mixing of $|x^2 - y^2>$ and $|3z^2 - r^2>$ orbital degrees of freedom is found to be strong and momentum dependent. With inclusion of $k_z$ dispersion, the 3D FS including its various pieces is reproduced in substantial detail. Moreover, our model accurately describes the delicate minority t$_{2g}$ FS pocket. 

Since the e$_g$ mixing has a pronounced effect on the shape of the FS, an accurate model allowing precise parameterization of the band structure is crucially important for modeling transport properties. Such a model would also provide a springboard for further theoretical work on strongly correlated electron systems, including Monte Carlo simulations to uncover the exciting many-body physics of the manganites\textsuperscript{30,31}. Moreover, a precise description of the minority t$_{2g}$ band is needed for the design of efficient spintronics devices. In this way, the TB models discussed in this study would also help develop the applications potential of the manganites.

\section{Methods}  
The first-principles calculations were done using the WIEN2K\textsuperscript{32,33} code. The electronic structure was calculated within the framework of the density-functional theory\textsuperscript{34,35} using linearized augmented plane-wave (LAPW) basis\textsuperscript{36}. Exchange-correlation effects were treated using the generalized gradient approximation (GGA)\textsuperscript{37}. A rigid band model was invoked for treating doping effects on the electronic structure along the lines of Ref. 25, but we expect our results to be insensitive to a more realistic treatment of doping effects using various approaches\textsuperscript{38-41}. We used muffin-tin radius ($R_{MT}$) of $1.80$ Bohr for both O and Mn, and $2.5$ Bohr for Sr and La. The integrals over the Brillouin zone were performed using a tetrahedron method with a uniform $14\times14\times14$ k-point grid. The ARPES experiments were performed on cleaved single crystals at beam lines 7.0.1 and 12.0.1 of the Advanced Light Source, Berkeley.

\section{References}
\begin{enumerate}
\item Salamon, M.B. \& Jaime, M. The physics of manganites: structure and transport. {\it Rev. Mod. Phys.} {\bf 73}, 583 (2001).
\item Montano, P.A. et al. Inelastic magnetic X-ray scattering from highly correlated electron systems: La$_{1.2}$Sr$_{1.8}$Mn$_{2}$O$_{7}$, La$_{0.7}$Sr$_{0.3}$MnO$_{3}$ and Fe$_{3}$O$_{4}$. {\it J. of Phys. and Chem. of Solids} {\bf 65}, 1999-2004 (2004).
\item von Helmolt, R., Wecker, J., Holzapfel, B., Schultz, L. \& Samwer, K. Giant negative magnetoresistance in perovskitelike La$_{2/3}$Ba$_{1/3}$MnO$_{x}$ ferromagnetic films. {\it Phys. Rev. Lett.} {\bf 71}, 2331 (1993).
\item de Groot, R.A., Mueller, F.M., van Engen, P.G. \& Buschow, K.H.J. New class of materials: half-metallic ferromagnets. {\it Phys. Rev. Lett.} {\bf 50}, 2024 (1983).
\item Oles, A.M., \& Feiner, L.F. Exchange interactions and anisotropic spin waves in bilayer manganites. {\it Phys. Rev. B} {\bf 67}, 092407 (2003).
\item Pickett, W.E. \& Moodera, J.S. Half metallic magnets. {\it Physics Today} {\bf 54}, 39 (2001).
\item Wolf, S.A. et al. Spintronics: a spin-based electronics vision for the future. {\it Science} {\bf 294}, 1488-1495 (2001)
\item Seshadri, R., Maignan, A., Hervieu, M., Nguyen, N. \& Raveau, B. Complex magnetotransport in LaSr$_{2}$Mn$_{2}$O$_{7}$. {\it Solid State Comm.} {\bf 101}, 453-457 (1997).
\item de Boer, P.K. \& de Groot, R.A. Electronic structure of the layered manganite LaSr$_{2}$Mn$_{2}$O$_{7}$. {\it Phys. Rev. B} {\bf 60}, 10758 (1999). 
\item Kubota, M. et al. Relation between crystal and magnetic structures of layered manganite La$_{2-2x}$Sr$_{1+2x}$Mn$_{2}$O$_{7}$ (0.30$\leq$x$\leq$0.50). {\it J. Phys. Soc. Jpn.} {\bf 69}, 1606-1609 (2000).
\item Sun, Z. et al. Electronic structure of the metallic ground state of La$_{2-2x}$Sr$_{1+2x}$Mn$_{2}$O$_{7}$ for $x\approx 0.59$ and comparison with $x=0.36,0.38$ compounds as revealed by angle-resolved photoemission. {\it Phys. Rev. B} {\bf 78}, 075101 (2008).
\item Barbiellini, B. et al. Extracting d-orbital occupancy from magnetic Compton scattering in bilayer manganites. {\it J. of Phys. and Chem. of Solids} {\bf 66}, 2197–2201 (2005).
\item Sun, Z. et al. Minority-spin t$_{2g}$ states and the degree of spin polarization in ferromagnetic metallic La$_{2-2x}$Sr$_{1+2x}$Mn$_{2}$O$_{7}$ $(x = 0.38)$. {\it Scientific Reports} {\bf 3}, 3167 (2013). 
\item Barbiellini, B., Moroni, E.G. \& Jarlborg, T. Effects of gradient corrections on electronic structure in metals. {\it J. Phys.: Condens. Matter} {\bf 2}, 7597 (1990).
\item Zener, C. Interaction between the d-Shells in the transition metals. II. ferromagnetic compounds of manganese with perovskite structure. {\it Phys. Rev.} {\bf 82}, 403 (1951).
\item Barbiellini, B. et al. Role of oxygen electrons in the metal-insulator transition in the magnetoresistive oxide La$_{2-2x}$Sr$_{1+2x}$Mn$_{2}$O$_{7}$ probed by compton scattering. {\it Phys. Rev. Lett.} {\bf 102}, 206402 (2009).
\item Ederer, C., Lin, C., \& Millis, A.J. Structural distortions and model Hamiltonian parameters: From LSDA to a tight-binding description of LaMnO$_{3}$. {\it Phys. Rev. B} {\bf 76}, 155105 (2007).
\item Saniz, R., Norman, M.R. \& Freeman, A.J. Orbital mixing and nesting in the bilayer manganites 
La$_{2-2x}$Sr$_{1+2x}$Mn$_{2}$O$_{7}$. {\it Phys. Rev. Lett.} {\bf 101}, 236402 (2008).
\item Sun, Z. et al. Quasiparticlelike peaks, kinks, and electron-phonon coupling at the $(\pi,0)$ regions in the CMR oxide La$_{2-2x}$Sr$_{1+2x}$Mn$_{2}$O$_{7}$. {\it Phys. Rev. Lett.} {\bf 97}, 056401 (2006). 
\item Slater, J.C. \& Koster, G.F. Simplified LCAO method for the periodic potential problem. {\it Phys. Rev.} {\bf 94}, 1498 (1954).
\item Sahrakorpi, S., Lindroos, M., Markiewicz, R.S. \& Bansil, A. Evolution of midgap states and residual three dimensionality in La$_{2-x}$Sr$_{x}$CuO$_{4}$. {\it Phys. Rev. Lett.} {\bf 95}, 157601 (2005).
\item Bansil, A., Lindroos, M., Sahrakorpi, S., \& Markiewicz, R.S. Influence of the third dimension of quasi-two-dimensional cuprate superconductors on angle-resolved photoemission spectra. {\it Phys. Rev. B} {\bf 71}, 012503 (2005).
\item Campuzano, J.C., Smedskjaer,  L.C., Benedek, R., Jennings, G., \& Bansil, A. Fermi surface in YBa$_2$Cu$_3$O$_6.9$: evidence from ARPES and positron 2D-ACAR. {\it Phys. Rev. B} {\bf 43}, 2788 (1991).
\item Smedskjaer, L.C., Bansil, A., Welp, U., Fang, Y., \& Bailey, K. G. Positron studies of metallic YBa$_2$Cu$_3$O$_{7-x}$. {\it J. Phys. Chem. Solids} {\bf 52}, 1541 (1991).
\item Mijnarends, P.E. et al. Magnetic momentum density, fermi surface, and directional magnetic compton profiles in LaSr$_{2}$Mn$_{2}$O$_{7}$ and La$_{1.2}$Sr$_{1.8}$Mn$_{2}$O$_{7}$. {\it Phys. Rev. B} {\bf 75}, 014428 (2007).
\item Anderson, P.W. \& Hasegawa, H. Considerations on double exchange. {\it Phys. Rev.} {\bf 100}, 675 (1955).
\item Sun, Z. et al. Nonmonotonic fermi surface evolution and its correlation with stripe ordering in bilayer manganites. {\it Phys. Rev. B} {\bf 86}, 201103(R) (2012).
\item Li, Y. et al. Temperature-dependent orbital degree of freedom of a bilayer manganite by magnetic compton scattering. {\it Phys. Rev. Lett.} {\bf 93}, 207206 (2004).
\item Koizumi, A. et al. Study of the e$_{g}$ orbitals in the bilayer manganite La$_{2-2x}$Sr$_{1+2x}$Mn$_{2}$O$_{7}$ by using magnetic compton-profile measurement. {\it Phys. Rev. Lett.} {\bf 86}, 5589 (2001)
\item Salafranca, J., Alvarez, G. \& Dagotto, E. Electron-lattice coupling and partial nesting as the origin of fermi arcs in manganites. {\it Phys. Rev. B} {\bf 80}, 155133 (2009).
\item Monkman, E.J. et al. Quantum many-body interactions in digital oxide superlattices. {\it Nature Materials} {\bf 11}, 855-859 (2012).
\item Schwarz, K. \& Blaha, P. Solid state calculations using WIEN2k. {\it Comput. Mater. Sci.} {\bf 28}, 259-273 (2003).
\item Blaha, P., Schwarz, K., Sorantin, P. \& Trickey, S.B. Full-potential, linearized augmented plane wave programs for crystalline systems. {\it Comput. Phys. Commun.} {\bf 59}, 399-415 (1990).
\item Hohenberg, P. \& Kohn, W. Inhomogeneous electron gas. {\it Phys. Rev.} {\bf 136}, B864 (1964).
\item Kohn, W. \& Sham, L.J. Self-consistent equations including exchange and correlation effects. {\it Phys. Rev.} {\bf 140}, A1133 (1965).
\item Andersen, O.K. Linear methods in band theory. {\it Phys. Rev. B} {\bf 12}, 3060 (1975).
\item Perdew, J.P., Burke, K. \& Ernzerhof, M. Generalized gradient approximation made simple. {\it Phys. Rev. Lett.} {\bf 77}, 3865 (1996). 
\item Bansil, A., Rao, R.S., Mijnarends, P.E. \& Schwartz, L. Electron momentum densities in disordered muffin-tin alloys. {\it Phys. Rev. B} {\bf 23}, 3608 (1981).
\item Lin, H., Sahrakorpi, S., Markiewicz, R.S., \& Bansil, A. Raising Bi-O bands above the Fermi energy level of hole doped Bi2212 and other cuprate superconductors. {\it Phys. Rev. Lett.} {\bf 96}, 097001 (2006).
\item Asonen, H. et al. Angle-resolved photoemission study of (100), (110), and (111) surfaces of Cu$_{90}$Al$_{10}$. {\it Phys. Rev. B} {\bf 25}, 7075 (1982).
\item Khanna, S.N., Ibrahim, A.K., McKnight, S.W., \& Bansil, A. d-band filling in Ni-P metallic glasses. {\it Solid State Commun.} {\bf 55}, 223 (1985).
\end{enumerate}

\section{Acknowledgements}
This research was performed while one of us (M.B.) was on sabbatical leave
from Boston University. The work at Northestern University was supported by the US Department of Energy, Office of
Science, Basic Energy Sciences contract DE-FG02-07ER46352, and
benefited from the allocation of supercomputer time at NERSC through DOE grant number DE-AC02-05CH11231, and
Northeastern University's Advanced Scientific Computation Center
(ASCC). We thank H. Zheng and J. Mitchell for providing the crystals, and J.F. Douglas, A. Fedorov, E. Rotenberg, and Q. Wang for help with the experiments. The work at the University of Colorado Boulder was supported by the US Department of Energy under grant number DE-FG02-03ER46066.  The Advanced Light Source is supported by the Director, Office of Science, Office of Basic Energy Sciences, of the U.S. Department of Energy under Contract No. DE-AC02-05CH11231.

\section{Author Contributions}
M.B., H.L., C.L., H.H., R.S.M., B.B., Z.S., D.S.D., and A.B. all contributed to the research reported in this study and the writing of the manuscript.

\section{Additional Information}
The authors declare no competing financial interests.

\begin{figure}[p]
\begin{center}
  \includegraphics[scale=.7]{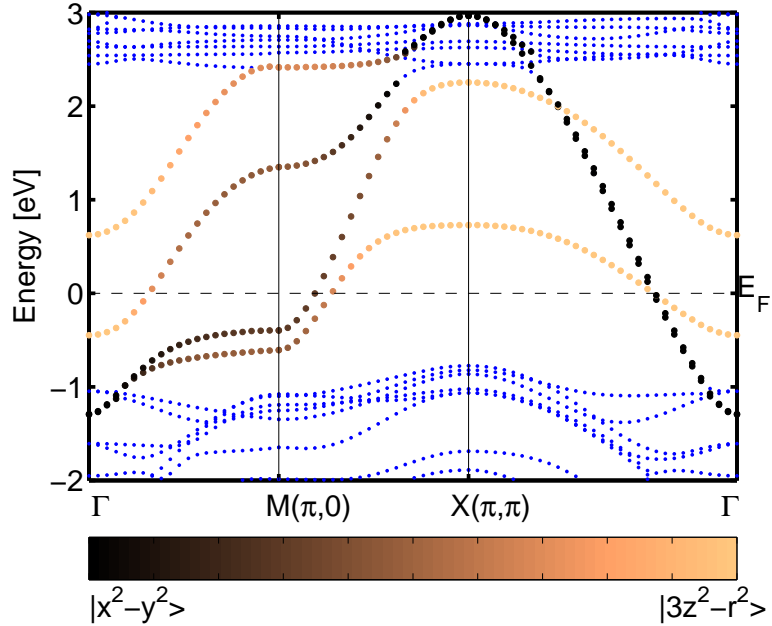}
\caption{(color online) Majority-spin band structure at $k_z=0$ in the FM state for $x=0.50$. The color scale identifies the $|x^2 - y^2>$ and $|3z^2 - r^2>$ characters of various bands.}
\end{center}
\end{figure}

\begin{figure}[p]
\begin{center}
  \includegraphics[width=.7\textwidth]{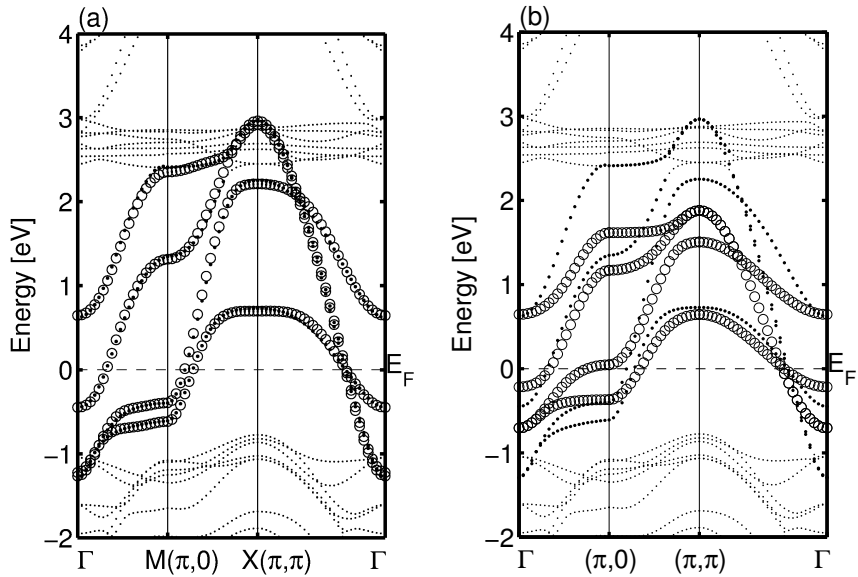}
\caption{
  (a) Majority-spin TB band structure (open circles) obtained from the 2D model discussed in the text is superimposed on the corresponding first-principles bands in the FM state for $x=0.50$ (solid dots); (b) Same as (a), except that the TB bands here (open circles) are based on a simple two parameter model.}
\end{center}
\end{figure}

\begin{figure}[p]
\begin{center}
  \includegraphics[scale=.7]{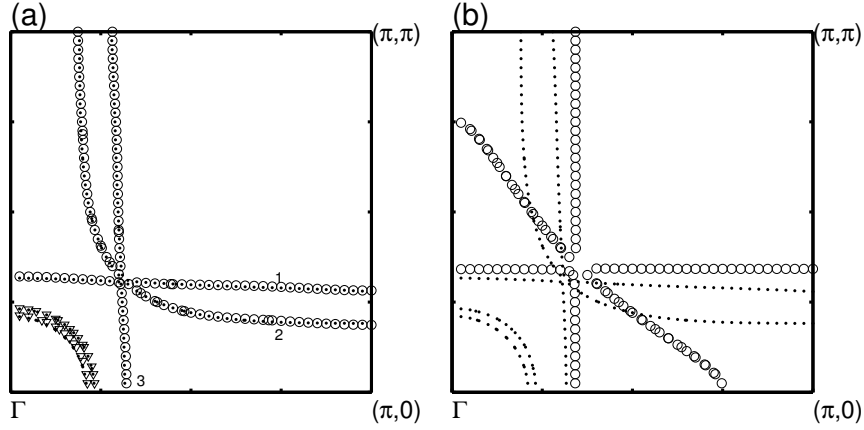}
\caption{ 
(a) TB majority-spin Fermi surface based on the bands of Figure 2a(open circles), and the minority-spin Fermi surface from bands of Figure 4 (triangles) are superposed on the first-principles results for $x=0.50$ in the FM state (solid dots). (b) Same as (a), except that the TB bands here (open circles) are based on a simple two parameter model.
}
\end{center}
\end{figure}

\begin{figure}[p]
\begin{center}
  \includegraphics[scale=.7]{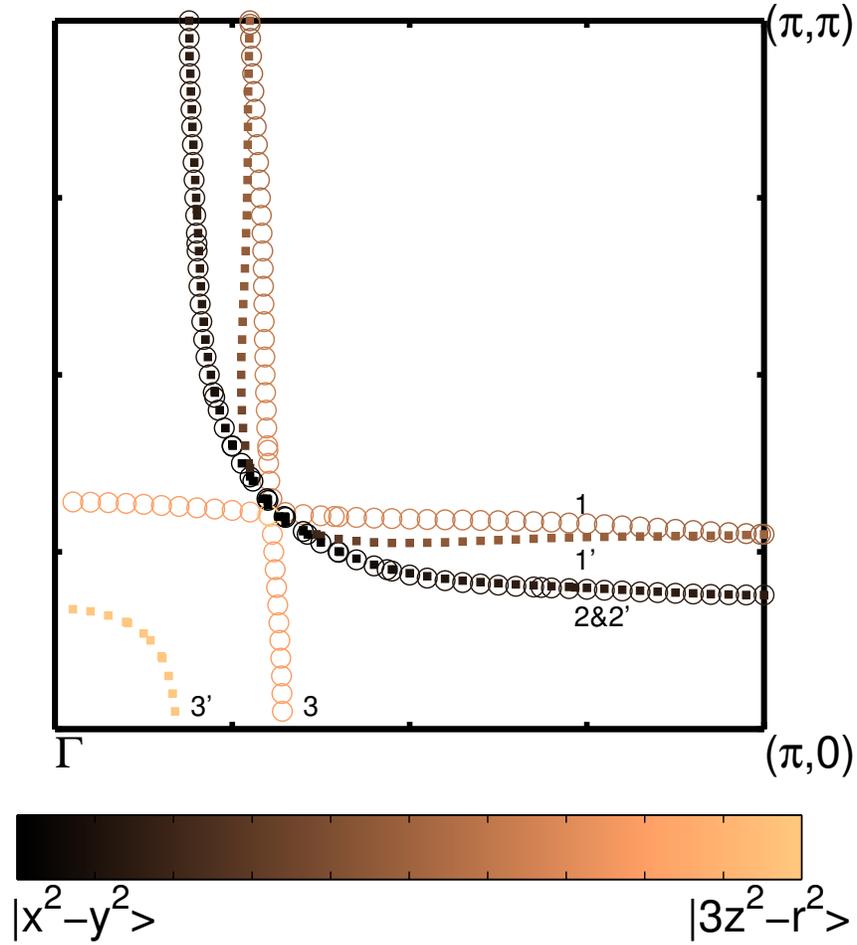}
\caption{(color online)   Majority-spin Fermi surfaces at $k_zc=0$ (circles) and $k_zc=2\pi$ (square) based on the 3D TB model in the FM state for $x=0.50$, as discussed in the text. The color scale identifies the $|x^2 - y^2>$ and $|3z^2 - r^2>$ characters of various bands.}
\end{center}
\end{figure}

\begin{figure}[p]
\begin{center}
  \includegraphics[scale=.7]{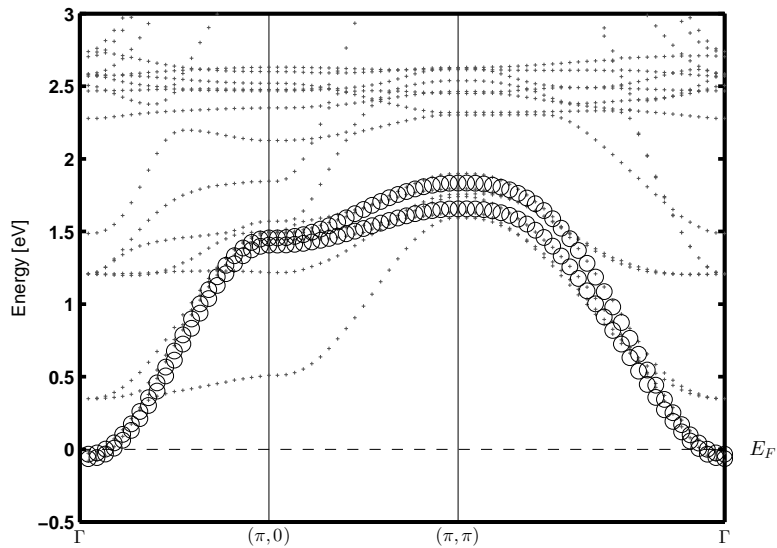}
\caption{ Minority-spin band structure obtained from the TB 2D model (open circles) discussed in the text is superimposed on the first-principles results at $k_z=0$ in the FM state for $x=0.50$.(solid dots).}
\end{center}
\end{figure}

\begin{figure}[p]
\begin{center}
  \includegraphics[scale=.56]{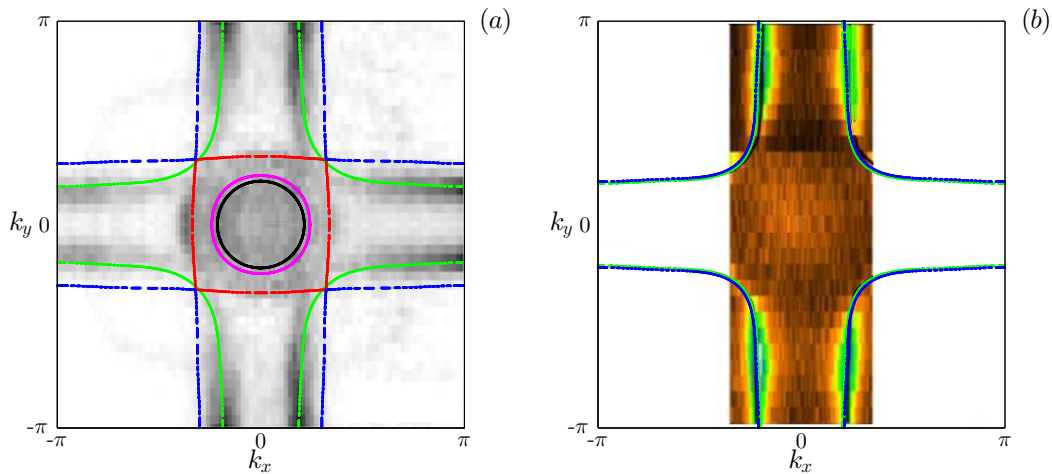}
\caption{ (color online) (a) Experimental Fermi surfaces of bonding bands for $x=0.38$\textsuperscript{13} (FM) are overlaid with the results of the 2D TB model. Theoretical Fermi surfaces are derived from various orbitals as follows: $|3z^2 - r^2>$ (red line); bonding $|x^2 - y^2>$ (blue line), which is not seen in experiments due presumably to effects of the ARPES matrix element at this photon energy; anti-bonding $|x^2 - y^2>$ (green line); and, the minority pockets (pink and black lines). (b) Experimental Fermi surface at $x=0.59$\textsuperscript{11} (AFM) overlaid with the 2D TB model predictions. Theoretical Fermi surfaces derived from the bonding and anti-bonding $|x^2 - y^2>$ bands are shown by blue and green lines, respectively.}
\end{center}
\end{figure}

\begin{table}[p]
\caption{Values of various tight-binding parameters for the majority spin bands (in meV).}
\begin{tabular}{||c|r|r||c|r|r||}
\hline\hline
 &2-D model&3-D model&  &2-D model&3-D model \\
\hline\hline
$t_{11}$    	&-669	&-670	&$E_z$     			&-305	&-337  \\     \hline
$t_{22}$    	&-678	&-649	&$t'_{11}$  			& 149	& 153  \\     \hline
$t_{12}$    	&-579	&-575	&$t'_{22}$    		&-299	&-300  \\     \hline
$t_{bi2}$	&-652	&-588	&$t'_{bi2}$			& 105	& 176  \\     \hline \hline
$t''_{11}$  	&-123	&-127	&$t'''_{11}$			& -28	& -28  \\     \hline
$t'_{12}$   	&-30 	&-19		&$t''_{12}$   		& -32	& -36  \\     \hline
$t_{bi1}$	&-22 	&12		&$\mu_{\uparrow}$   	& 920	& 912\\     \hline \hline
$t_z$    	& -  	&-126	&$t'_z$    			& -  	& 25   \\     \hline \hline
\end{tabular}
\end{table}

\begin{table}[p]
\caption{Values of various tight-binding parameters for the minority-spin bands (in meV).}
\begin{tabular}{||c|r||c|r||}
\hline\hline
$t_{11}$   		&-466.6	&$t_{22}$    			&-430.1  \\     \hline
$t'_{11}$   		&-277.6	&$t'_{22}$    			&-305.2 \\     \hline
$t''_{11}$    	&-2		&$t''_{22}$    			&24.7   \\     \hline
$t'''_{11}$    	&-8.7	&$t'''_{22}$    			&-24    \\     \hline \hline
$\Delta$			    	&114.9	&$\mu_{\downarrow}$ 	&1189.16  \\     \hline \hline
\end{tabular}
\end{table}

\begin{table}[p]
\caption{Values of various parameters obtained for our 2D and 3D models over the doping range $0.38$-$0.59$ (all per Mn atom): change in E$_F$ ($\Delta E_{F}$); number of minority electrons ($\Delta n$); number of majority electrons ($1-x-\Delta n$); total number of electrons ($1-x$); magnetic moment ($\mu_{B}$); and the canting angle ($\theta_{cant}$).}
\begin{tabular}{|c||c|c|c|c|c|c|c||}
\hline\hline
&$x$&$\Delta E_{F}$ $(eV)$&$\Delta n$&$1-x-\Delta n$&$1-x$&Moment $(\mu_{B})$&$\theta (^{\circ})$\\\hline\hline
2D&0.38	&0.173	&0.037	&0.582	&0.62	&3.544	&0.0\\
&0.40	&0.129	&0.040	&0.559	&0.60	&3.519	&6.3\\
&0.45	&0.082	&0.029	&0.520	&0.55	&3.491	&63\\
&0.48	&0.150	&0.034	&0.487	&0.52	&3.455	&180\\
&0.50	&0.111	&0.036	&0.463	&0.50	&3.427	&180\\
&0.59	&-0.039 &0.000	&0.410	&0.41	&3.410	&180\\\hline\hline
3D&0.38	&0.167	&0.044	&0.575	&0.62	&3.531	&0.0\\
&0.40	&0.140	&0.037	&0.562	&0.60	&3.525	&6.3\\
&0.45	&0.090	&0.026	&0.523	&0.55	&3.497	&63\\
&0.48	&0.137	&0.037	&0.482	&0.52	&3.445	&180\\
&0.50	&0.107	&0.030	&0.469	&0.50	&3.439	&180\\
&0.59	&-0.048 &0.000	&0.410	&0.41	&3.410	&180\\
\hline\hline
\end{tabular}
\end{table}

\begin{table}[p]
\caption{ Values of various parameters obtained for our 2D and 3D models for the FM state over the doping range $0.38$-$0.50$ (all per Mn atom): change in E$_F$ ($\Delta E_{F}$); number of minority electrons ($\Delta n$); number of majority electrons ($1-x-\Delta n$); total number of electrons ($1-x$); and, the magnetic moment ($\mu_{B}$).}
\begin{tabular}{|c||c|c|c|c|c|c|c||}
\hline\hline
&$x$&$\Delta E_{F}$ $(eV)$&$\Delta n$&$1-x-\Delta n$&$1-x$&Moment $(\mu_{B})$\\\hline\hline
2D&0.38	&0.173	&0.037	&0.582	&0.62	&3.544\\
&0.40	&0.145	&0.031	&0.568	&0.60	&3.537\\
&0.45	&0.097	&0.020	&0.529	&0.55	&3.509\\
&0.48	&0.150	&0.032	&0.487	&0.52	&3.455\\
&0.50	&0.000	&0.001	&0.498	&0.50	&3.497\\\hline\hline
3D&0.38	&0.167	&0.044	&0.575	&0.62	&3.531\\
&0.40	&0.140	&0.037	&0.562	&0.60	&3.524\\
&0.45	&0.074	&0.015	&0.535	&0.55	&3.520\\
&0.48	&0.031	&0.005	&0.514	&0.52	&3.508\\
&0.50	&0.000	&0.006	&0.493	&0.50	&3.487\\
\hline\hline
\end{tabular}
\end{table}

\end{document}